\documentstyle[12pt]{article}

\begin{document}

\begin{titlepage}
\null\vspace{-62pt}

\pagestyle{empty}
\begin{center}

\vspace{1.0truein} {\Large\bf Gravitational effects on critical
                       Q-balls }

\vspace{1in}
{\large Dimitrios Metaxas} \footnote{previous address: Demokritos Institute for Nuclear Physics, Athens, Greece.} \\
\vskip .4in
{\it Department of Physics,\\
National Technical University of Athens,\\
Zografou Campus, 15780 Greece\\
metaxas@central.ntua.gr}\\

\vspace{0.5in}

\vspace{.5in}
\centerline{\bf Abstract}

\baselineskip 18pt
\end{center}

In a cosmological phase transition in theories that admit Q-balls
there is a value of the soliton charge above which the soliton
becomes unstable and expands, converting space to the true vacuum,
much like a critical bubble in the case of ordinary tunneling. Here
I consider the effects of gravity on these solitons and I calculate
the lowest gravitational corrections to the critical radius and
charge.

\end{titlepage}

\newpage
\pagestyle{plain}
\setcounter{page}{1}
\newpage

Non-topological solitons of the Q-ball type, exist in field theories
with a global $U(1)$ symmetry \cite{Coleman, Lee-Pang}, may be
formed in the early Universe with various mechanisms, and have
important cosmological implications \cite{kuz1, enq, kas, postma}.

 Here I will be interested in the case of Q-balls that are formed
in the context of a cosmological phase transition and may act as
seeds for its nucleation. In the case of a first-order phase
transition, when the symmetric vacuum becomes metastable, the
potential energy density in the interior of the Q-balls becomes
negative with respect to the potential energy density of the
symmetric vacuum. Then, if a Q-ball has  large enough charge, it
becomes unstable and expands, converting space  to  the true vacuum,
much  like a critical bubble in the case of ordinary tunneling
\cite{Spector, Kusenko, Ellis, metaxas}. Q-ball stability has also
been investigated in other models that may be of cosmological
relevance \cite{minos, copeland}. These results may be important for
cosmological phase transitions in theories that admit
non-topological solitons, such as the Minimal Supersymmetric
Standard Model (MSSM) \cite{Kusenko, enq, Ellis}.

 In these models,
that exhibit a phase transition at or above the TeV scale, the
calculation of the tunneling rate and the kinetics of the phase
transition is done along the lines of \cite{linde} and gives a
specific picture of the phase transition, its progress and its
completion. However, when the theory admits topological or
non-topological defects (solitons, monopoles, or Q-balls as in MSSM)
the picture of the phase transition may be quite different
\cite{Coleman, Spector, Kusenko, metaxas}. These defects may act as
nucleations sites for the phase transition and accelerate its
progress, facilitating its completion.

In the case of Q-balls, we are interested in the critical soliton
charge $Q_c$ (and the corresponding critical soliton radius $R_c$)
above which the solitons become unstable and expand, converting
space to the true vacuum and accelerating the phase transition
\cite{Coleman, Spector}.

 Here I will calculate the
first-order gravitational corrections to this effect, that can
mediate tunneling in curved space-time \cite{CdL}. The effects of
gravity to the ordinary bubble nucleation rate can be very distinct
\cite{ejw, balek, klee}, due mainly to the finiteness of the
compactified de Sitter space. Here, however, I will be interested
only in the case where the relevant solitons are much smaller than
the radius of the background de Sitter space, and I will calculate
the lowest gravitational corrections to these solitons.

 In the
 case of a de Sitter background, the effect of the cosmological
 expansion that I derive here is to decrease the values of $Q_c$ and
 $R_c$, as is expected, thereby enhancing further the progress of the phase transition.
 Although the
effect is of higher order, the correction to $Q_c$ is finite and may
be important, especially for small values of the supercooling
parameter $\eta=(T_c-T)/T_c$, as will be shown here.

I  will use the gravitational equations that describe Q-balls in
curved space-time \cite{Multamaki} in the thin-wall approximation
and I will calculate the gravitational effects to the critical
radius and critical charge that these Q-balls must have in order to
mediate the phase transition.

Consider a quantum field theory
of a complex scalar field $\Phi$
with a $U(1)$ symmetric scalar potential
$U(\Phi)$.
For spherically symmetric configurations
with metric
\begin{equation}
ds^2 = B\,dt^2 - A\,dr^2 - r^2 \, d\Omega
\end{equation}
the action is
\begin{equation}
S = \int d^4 x \left\{ -\frac{\sqrt{A B}}{16 \pi G}
    + \sqrt{A B} \left[ g^{\mu \nu}
                        \partial_{\mu}\Phi \partial_{\nu}\Phi^{*}
                        - U(\Phi)
                       \right]
                       \right\}
\end{equation}
and for solutions of the Q-ball type
\begin{equation}
\Phi = e^{i \omega t} \, \phi(r)
\end{equation}
the field equations are
\begin{equation}
\left( \frac{r}{A} \right)^{'}
=
1-8\pi G r^2 \rho
\end{equation}
\begin{equation}
\frac{B^{'}}{B} + \frac{A^{'}}{A}
=
8 \pi G A r (\rho +p)
\end{equation}
\begin{equation}
\phi^{''} + \phi^{'} \frac{1}{r^2} \sqrt{\frac{A}{B}}
            \frac{\partial}{\partial r}
            \left( r^2 \sqrt{\frac{B}{A}} \right)
           - \frac{1}{2} A \frac{\partial U_{\omega}}{\partial\phi}
= 0
\end{equation}
where
\begin{equation}
U_{\omega}(\phi) = U(\phi)
                   - \frac{\omega^2}{B} \phi^2
\end{equation}
\begin{equation}
\rho = \frac{\omega^2}{B} \phi^2
      + \frac{1}{A}{\phi^{'}}^2 + U
\end{equation}
\begin{equation}
p = \frac{\omega^2}{B} \phi^2
      + \frac{1}{A}{\phi^{'}}^2 - U
\end{equation}.

Now we consider a potential that has a local
minimum $U_0$ at $\phi =0$
and a global minimum $U_c$ at $\phi = \phi_c$,
with the height of the barrier much greater than
$\varepsilon = U_0 -U_c$
the difference between the two minima.
Like in the flat case \cite{Coleman, Spector}
this theory admits non-topological thin-wall solitons
of the Q-ball type \cite{Multamaki}, where
the field forms a bubble of radius $R$,
inside the bubble the field is essentially constant
and outside zero.
For this kind of potential there is no minimum $\omega$
since $\phi=0$ is already metastable,
so the solutions have $\omega \approx 0$
and inside the bubble $\phi \approx \phi_c$
(these correspond to the type II solitons of \cite{Multamaki}).

For the configuration inside the bubble
we make the ansatz
\begin{equation}
A = \frac{1}{1- \alpha r^2}\,\,,\,\,\,\,\,
B = 1-\beta r^2
\end{equation}
where, in the weak gravity approximation,
\begin{equation}
\alpha R^2 \,\,,\,\,\,\, \beta R^2\, << 1
\end{equation}
and we get
\begin{equation}
\alpha = \frac{8 \pi G}{3}
         (\omega^2 \phi_c^2 + U_c)
\end{equation}
\begin{equation}
\beta = \frac{8 \pi G}{3}
        (-2\omega^2 \phi_c^2 + U_c)
\end{equation}
and our approximation is self-consistent when
\begin{equation}
G \omega^2 \phi_c^2 R^2 \,,\,\,
G U_c R^2 \, << 1
\label{approx1}
\end{equation}.

Now we calculate the values of the charge $Q$ and the energy $E_Q$
of this configuration, making the further approximation $\omega^2
\phi^2 << U_c$, which holds for this kind of potentials. So our
final approximations, to be checked later for self-consistency, are
\begin{equation}
G \omega^2 \phi_c^2 R^2 \, <<\, G U_c R^2 \, <<1.
 \label{approx1}
\end{equation}
In our further calculations we keep
the terms that are of lowest order
in $\omega$ and $G$.

In this approximation the expression for the total charge gives
\begin{equation}
Q=8\pi\omega\int_0^R dr\, r^2
   \sqrt{AB} \phi_c^2 \approx 8\pi\omega\phi_c^2
   \frac{R^3}{3}
\end{equation}
and the expression for the soliton energy $ E_Q = \int d^3 x \rho $
becomes
\begin{equation}
E_Q = E_{\omega} + E_{surf} + E_{vol}
\end{equation}
with
\begin{equation}
E_{\omega} = 4\pi \omega^2 \phi_c^2
            (\frac{1}{3}R^3 + \frac{1}{5}\gamma R^5)
\end{equation}
where
\begin{equation}
\gamma=\frac{8\pi G}{3} U_c
\end{equation}
and
\begin{equation}
E_{vol} = -\frac{4\pi R^3}{3} \varepsilon.
\end{equation}
For the surface term, we assume that the soliton surface has a
thickness $\delta$ and an average value of the potential $U_s$.
Minimizing with respect to $\delta$ \cite{Spector} we get
\begin{equation}
E_{surf}=4 \pi R^2 \phi_c \sqrt{U_s}(2-\gamma R^2).
\end{equation}
Now the soliton radius can be found minimizing the energy with
respect to $R$, and the critical radius and charge for these
solitons are found from the simultaneous solution of \cite{Spector,
Kusenko}
\begin{equation}
\frac{dE_Q}{dR}=\frac{d^2 E_Q}{dR^2}=0.
\end{equation}
The critical radius in this approximation turns out to be
\begin{equation}
R_c=\frac{10\phi_c\sqrt{U_s}}{3 \varepsilon
        \left( 1+
        \frac{52 \cdot 10^2}{5 \cdot 3^4}
         \frac{8 \pi G}{3}
         \frac{U_c U_s \phi_c^2}{\varepsilon^2}\right)}
\end{equation}
and for the critical charge we get
\begin{equation}
Q_c^2 =\frac{2^{12}\pi^2 5^5}{3^9}
       \frac{\phi_c^8 U_s^3}{\varepsilon^3 \left( \varepsilon^2
         +\frac{17 \cdot 2^5 \cdot 5^2 \pi G}{3^4} \phi_c^2 U_c U_s
           \right)}.
 \label{result}
\end{equation}
These are the first order gravitational corrections to the flat
space expressions (that we get for $G=0$). We see that the effect of
the cosmological expansion has been to lower the values of the
critical charge and the critical radius, as is expected, thereby
facilitating the phase transition.

Now, in order to check for the validity of our approximations, we
take $U_s \sim m^2 \phi_c^2$ where $m$ is a typical mass scale of
the theory, and we see that our approximations (\ref{approx1}) are
equivalent to
\begin{equation}
\frac{m \phi_c^2 \sqrt{U_c}}{M_P} << \varepsilon << U_c
 \label{con1}
\end{equation}
where $M_P=G^{-1/2}$ is the Planck mass. So our approximations for
this problem are self-consistent. Although the scale of the specific
model (MSSM or other) is fixed, of order TeV or higher, the value of
$U_c$, which corresponds to the cosmological constant term, is
arbitrary, provided it it satisfies (\ref{con1}).

It is easy to see that the condition for stability, $E_Q < Q m$, is
satisfied for generic values of the potential well below the Planck
scale. We can also check that for the values of the parameters that
satisfy (\ref{con1}) the soliton radius is much smaller than the
radius of the compactified de Sitter space $R_{dS}\sim
M_P/\sqrt{U_c}$, so we are indeed in the limit of weak gravity.

However, even though we are in the limit of weak gravity, the
correction to $Q_c$ derived here, may be finite, and even large for
the cosmological phase transition. The production rate for
non-topological solitons with charge $Q$ in a cosmological setting
is generically proportional to $e^{-Q}$ \cite{Kusenko, metaxas} and
a change in $Q_c$ gives important corrections regarding the progress
and completion of the phase transition.

In order to see this we can consider a generic effective potential
 $U(\phi, T)$,
depending on the temperature $T$, with the two degenerate minima (at
$\phi=0$ and $\phi=\phi_c$)
 at the critical temperature $T_c$. Then, slightly below
the critical temperature, we have \cite{linde}
\begin{equation}
\varepsilon= g \,T^2 \, \phi_c^2 \, \eta
\end{equation}
with $g$ a coupling constant of the model  and $\eta$ the
supercooling parameter $\eta=(T_c-T)/T$.
 Then we can calculate from (\ref{result}) the decrease of the
critical charge due to gravity
\begin{equation}
\delta Q_c \sim \frac{\phi_c^2}{m^2 \eta^{5/2}}
                \left( \frac{m^2\phi_c^4\,U_c}{M_P^2\,\varepsilon^2}
                \right).
 \label{res2}
 \end{equation}
We see that, although our approximation (\ref{con1}) demands the
second term in (\ref{res2}) to be small, the result for $\delta Q_c$
may be large for sufficiently small values of the supercooling
parameter $\eta$. It is for these cases of a "fast" phase transition
that the results presented here are relevant and may give important
corrections to the usual calculation of the false vacuum decay rate.

Generally the presence of solitonic configurations (of the type
described here or other) greatly facilitates the progress and
completion of the phase transition since they act as seeds for its
nucleation, however the actual dynamics and kinematics of the
transition are quite complicated since they depend on the
interactions of these solitons with the heat bath. For a phase
transition that occurs at a scale $\mu$, we see from the previous
expressions that the corrections calculated here are important when
the transition proceeds with a  small value of supercooling $\eta
\sim \mu/M_P$. For the lowest scale supersymmetric models that admit
such configurations, with $\mu$ of order of a few TeV this
corresponds to $\eta\sim 10^{-16}$, that is very close to a second
order phase transition. This is the case when these solitons are
formed at high temperature, by thermal and charge fluctuations,
before the onset of the first order phase transition, and as soon as
the symmetric vacuum becomes metastable they become unstable and
expand converting the space in the true vacuum.
 Similar estimates hold when we are
interested in an intermediate scale phase transition, provided
(\ref{con1}) is satisfied, the actual dynamics of the transition,
however, may change these estimates.

In summary we see that the corrections calculated here have the
expected effects for the case
 of non-topological solitons in an expanding de Sitter background,
  that is to decrease the critical radius and critical charge of the
solitons. Although I worked in the limit of weak gravity, these
effects show that, in the general case of vacuum decay in curved
spacetime, such solitonic configurations may be important in the
calculation of the rate of the phase transition.

\vspace{0.5in} \centerline{\bf Acknowledgements} \noindent
 Parts of this work were done while visiting Demokritos Institute of
 Nuclear Physics and the National Technical University of Athens.
 I would like to thank the people of the Physics Departments for
 their hospitality.

\end{document}